\documentclass[%
 reprint,
 amsmath,amssymb,
 aps,nofootinbib,   
]{revtex4-1}
\usepackage{bm}
\PassOptionsToPackage{linktocpage}{hyperref}
\usepackage[hyperindex,breaklinks]{hyperref}
\usepackage{enumitem}
\usepackage{slashed}

\renewcommand{\theta}{\vartheta}

\renewcommand{\vec}[1]{\ensuremath{\boldsymbol{#1}}}

\newcommand{\bra}[1]{\ensuremath{\left< #1\,\right|}}
\newcommand{\ket}[1]{\ensuremath{\left|\, #1\right>}}

\usepackage{array}
\usepackage{mathtools}

\usepackage{etoolbox}

\begin{document} 

\title{Entropy Constraints on High Spin Particles}

\author{Markus Dierigl$^{1,2}$ and Gia Dvali$^{3,4,5}$}
\affiliation{%
$^1$
Institute for Theoretical Physics, Utrecht University, 3584 CC Utrecht, The Netherlands,
}%
\affiliation{%
$^2$
Institute of Physics, University of Amsterdam, 1098 XH Amsterdam, The Netherlands,
}%
\affiliation{%
$^3$ 
Arnold Sommerfeld Center, Ludwig-Maximilians-Universit\"at, Theresienstra{\ss}e 37, 80333 M\"unchen, Germany, 
}%
 \affiliation{%
$^4$ 
Max-Planck-Institut f\"ur Physik, F\"ohringer Ring 6, 80805 M\"unchen, Germany
}%
 \affiliation{%
$^5$ 
Center for Cosmology and Particle Physics, Department of Physics, New York University, 726 Broadway, New York, NY 10003, USA
}%

\date{\today}

\begin{abstract}
\noindent
Elementary particles of large spin $s$ store quantum information in degenerate states and therefore are subject to the Bekenstein entropy bound. We observe that for sufficiently large $s$ the bound is violated unless the particle acquires a new associated length-scale different from its Compton wavelength. This can be regarded as a glimpse of stringiness. Moreover, this bound is independent of gravity. The inclusion of gravity additionally generates a new scale at which the thermality of the black hole radiation is violated by the emission of a high spin particle. This bound can be understood as the black hole species bound, i.e.\ an induced quantum gravity cutoff-scale given by $M_P/\sqrt{s}$. The two bounds carry qualitatively different information.  
\end{abstract}

\maketitle
\section{Non-gravitational bound}

\noindent
The Bekenstein entropy bound \cite{BekBound}, see also \cite{BrBound}, tells us that the maximal amount of information stored in a system of energy $M$ and radius $R$ is given by 
\begin{equation} \label{Bek1}
S_{\max} = 2 \pi MR \,. 
\end{equation}
This bound is independent of gravity and is expected to hold in any consistent theory, see \cite{History} for the historical development and a complete list of references. 

Recently \cite{Gia}, it was pointed out that the Bekenstein bound is directly tied to {\it unitarity}. Namely, an object of mass $M$ and size $R$ saturates the Bekenstein bound on the entropy exactly when the underlying theory saturates perturbative unitarity at the scale $1/R$. That is, when unitarity is saturated for processes with momentum transfer $\sim 1/R$. Correspondingly, the bound can be written in a form that directly captures the connection with unitarity, 
\begin{equation} \label{alpha}
S_{\max} = {1 \over \alpha} \,,
\end{equation}   
where $\alpha$ represents a dimensionless coupling constant evaluated at the energy scale $1/R$. The correlation between the saturation of the Bekenstein bound and unitarity is independent of the renormalizability of the theory. This has been checked explicitly for a list of objects such as solitons, baryons and instantons.

Even though the bound is mainly applied to constrain macroscopic objects, one-particle states have to satisfy it as well. In theories with space-time and internal symmetries an elementary particle of a given energy can appear in many degenerate states with respect to color, flavor and spin. Therefore, it must be assigned the corresponding entropy. The question is, what kind of implications the Bekenstein bound induces once it is applied to this entropy. This line of thought can have some interesting consequences. For example, in \cite{Gia} a possibility was pointed out, that for QCD-like theories with a large number of colors, confinement can be understood as a prevention mechanism against would-be violation of the Bekenstein entropy bound in the presence of asymptotic, colored particles. 
  
In the present note, we shall apply a similar reasoning to elementary particles of high spin. We will conclude that in order to prevent a violation of the Bekenstein bound, the elementary particles of very high spin must acquire a new associated length-scale. This length-scale must grow with spin and has to be larger than the Compton wavelength.
 
We would like to point out that the precise definition of the parameters $M$ and $R$ entering the bound has been a subject of investigation by several authors, see \cite{History} and references therein. Since there are no exactly localized quantum objects of finite energy, our application of the bound will be guided by physical reasonability and parametric control in connection with their relevant uncertainties.\\
  
Consider  a Poincar\'e-invariant theory that allows for asymptotic S-matrix states in form of an elementary particle of mass $M$ and spin $s$. The theory may include an effective UV-cutoff $\Lambda$, which is induced by gravity or some other physics, above which the usual notion of an elementary particle does not exist. For example, in gravity one cannot have elementary particles with masses way beyond the Planck scale $M_P$. For such objects the gravitational radius becomes macroscopic and they cannot be treated as elementary. In our discussion we shall stay in an effective theory well below this cutoff. In particular, we must assume  $M \ll \Lambda$. Our goal is to show that in such a case the particle cannot have an arbitrarily high spin without violating the Bekenstein bound. That is, the particles of high spin necessarily lose the property of being elementary in the usual sense and belong to the UV-theory above $\Lambda$. Since this is the case for all the known consistent cases with particles of high spin -- such as, e.g., string theory or QCD -- one may argue that we are pushing at an open door. Nevertheless, gaining a novel understanding of the subject from the point of view of the Bekenstein entropy bound can be of fundamental importance and is worth to point out.
 
In order to make our arguments clean we assume that the particle in question is stable. However, the argument is equally applicable to unstable particles provided their life-time is larger than the relevant time-scales in our thought experiment.\\
    
Our assumptions imply that a one-particle state $\ket{1}_{\vec{k},\epsilon}$, characterized by a definite momentum $\vec{k}$ and spin-polarization $\epsilon$, represents a well-defined state in the Hilbert space. This state is obtained by an action of a corresponding creation operator on the S-matrix vacuum, $\ket{1}_{\vec{k},\epsilon} = \hat{a}^{\dagger}_{\vec{k}, \epsilon} \ket{0}$. If so, we can consistently consider their superpositions and form a Gaussian wave-packet localized within some radius $R$. 
   
Due to the possibly complicated underlying structure of the theory, the mass parameter $M$ might of course depend on $s$ as well as on an arbitrary number of coupling constants via renormalization. However, we are not making any assumptions about this dependence. At the end of the day our ignorance is all summed up in a single effective physical quantity $M$.
   
Notice, that we are not trying to confine the particles in a cavity and/or to stabilize the state using some external device. Such a situation was analysed in details in e.g.\ \cite{Bousso}. Of course, the process of localization would interfere with the Bekenstein bound since the device necessarily contributes to the total energy of the system. Furthermore, it is impossible to estimate this contribution without knowing the interaction strength and other parameters. However, this does not affect our argument since our approach is different. Our starting point is, that the elementary particle of high spin $s$ represents a legitimate asymptotic S-matrix state. This implies that a one-particle momentum state exists and it directly follows that the Hilbert space must contain arbitrary superpositions thereof. We then choose the state vector to represent a Gaussian wave-packet of any spin-polarization $\epsilon$ at some initial time $t=0$ 
 \begin{equation} \label{Gauss} 
 \ket{1_{\epsilon}} = C \int d^d k \, {\rm e}^{-R^2 \vec{k}^2} \ket{1}_{\vec{k},\epsilon} \,,
 \end{equation}
 with normalization constant $C$ chosen such that $\langle 1_{\epsilon} | 1_{\epsilon} \rangle = 1$. Being a legitimate state in the Hilbert space, we require that \eqref{Gauss} obeys the Bekenstein entropy bound \eqref{Bek1}.\\
 
To have parametric control, we need to take $R$ somewhat larger than the Compton wavelength $1/M$ since the wave-packet spreads with the velocity $v = (MR)^{-1}$. The energy of the wave-packet is given by the expectation value of the Hamiltonian,  
\begin{equation} \label{En}  
\bra{1_{\epsilon}} \hat{H} \ket{1_{\epsilon}} = M + {\mathcal O}(1/(MR^2)) \,.
\end{equation}
Additional corrections due to cutoff-dependent terms are more suppressed and hence are negligible. The first order correction indicated above can be derived as follows. Since we consider an expectation value in a one-particle state, only the free part of the Hamiltonian $\hat{H} = \sum_{\vec{k},\epsilon} \omega_{\vec{k},\epsilon} \, \hat{a}^{\dagger}_{\vec{k},\epsilon} \hat{a}_{\vec{k},\epsilon}$ contributes. By Poincar\'e symmetry and the fact that we are way below the cutoff $\Lambda$, we have $\omega_{\vec{k},\epsilon} \simeq \sqrt{M^2 + \vec{k}^2}$. Taking into account that the momenta effectively contributing to the wave-packet satisfy $|\vec{k}| \lesssim 1/R$, gives \eqref{En}.
  
Consequently, the Bekenstein bound \eqref{Bek1} must be satisfied with the corresponding time and energy uncertainties taken into account. However, the actual entropy of the state \eqref{Gauss} due to the spin-degeneracy is $S = \ln (2s +1)$. Thus, we arrive at the following bound  
\begin{equation} \label{Sbound}
\ln (2s +1)  \leqslant  2 \pi M R \,.
\end{equation} 
If the particle is a legitimate elementary degree of freedom, the relevant quantum length-scale associated with it is its Compton wavelength $1/M$. In such a case, the right hand side of the above equation is a matter of choice. For arbitrary $M$ we can choose the width of the wave-packet $R > 1/M$ at will. Then, for sufficiently high $s$, it is impossible to satisfy the bound. 

We can make the fact that the bound is violated exact by taking the limit 
\begin{equation} \label{limit}
s \rightarrow  \infty \,, \quad 2 \pi MR = \sqrt{ \ln (2s +1)}  \,.    
\end{equation} 
In this limit the velocity of the spread of the wave-packet vanishes and the wave-packet becomes an energy eigenstate. There is no ambiguity about the definition of energy and a localization width and the Bekenstein bound can be evaluated exactly. We have,  
\begin{equation} \label{limit}
{S \over S_{\rm max}} = \sqrt{ \ln (2s +1)} \,  \rightarrow \, \infty \,,     
\end{equation} 
and the Bekenstein bound is violated.\\
 
Of course, for finite parameters the values of $s$ that violate the bound are exponentially large and hence are too large for being of any relevance for the particles of the Standard Model. However, we arrive at an interesting conclusion. An object of sufficiently high spin cannot be an elementary particle in the following sense. In order to satisfy the Bekenstein bound on information storage the object must acquire a notion of a new length-scale which grows with spin. Notice that this is precisely what is happening in known examples with arbitrarily high-spin resonances such as QCD and string theory. In these theories, the high-spin states represent highly excited string vibrations and the role of the new length-scale is played by the string length.  

\section{Role of interactions}

\noindent
In the above reasoning we have at no point used information about the strength of the interaction. Indeed, the only assumption we have made is that the particle in question is a legitimate asymptotic S-matrix state. This may create possible confusion, since all we have said must apply to a free theory. It would be an extremely powerful statement if we could rule out a free theory of a high-spin particle, based on violation of the Bekenstein bound \eqref{Bek1}. Unfortunately, we cannot obtain such a strong statement for the following reason. As already pointed out in \cite{Gia}, the information content stored in a quantum state of a non-interacting theory is sterile and therefore meaningless.

In the present case, in order to read out the information stored in a state of a high-spin particle, we need a probe that is able to distinguish among the different spin-states. Thus, the theory must be interacting, since it only makes sense to constrain the amount of information that can be, at least in principle, decoded. This simple fact gives a hint of why the Bekenstein bound must be tied to unitarity as argued in \cite{Gia}. Indeed, the unitarity is directly sensitive to the strength of the interaction and thus to the speed by which the information stored in the object can be decoded. Therefore, when we apply the Bekenstein bound to a high-spin particle we must keep in mind that some minimal strength of a spin-sensitive interaction that allows for a read-out of the information must be assumed. In a free theory this is not the case. The fact that in a free theory the true information bound can never be violated is clear by the reformulation of the bound given in \eqref{alpha}, which is more convenient in such a case. Since for a free theory $\alpha =0$ the bound \eqref{alpha} is never violated. 
 
\section{Comparing with gravity} 
 
\noindent
In the discussion above no assumption was made about the presence of gravity in the theory. The inclusion of gravity requires a separate detailed investigation that will not be given here. However, it is useful to compare the non-gravitational bound to one immediate constraint due to gravity, which we shall derive next. For that, it is enough to remember that different spin-polarizations from the point of view of black hole physics represent different species and are therefore subject to the black hole species bound \cite{species}. According to this constraint, the number $N$ of elementary particle species of the low energy effective theory and the quantum gravity cutoff length-scale $L_*$ are related as, 
\begin{equation}  \label{species}
L_* \gtrsim \sqrt{N} L_P \,, 
\end{equation}
where $L_P = 1/M_P$ is the Planck length. Notice that in $N$ all the polarization-eigenstates count separately, as they all contribute equally into the black hole evaporation.
 
A simple argument for obtaining an analogous bound on the spin is to consider the thermal evaporation process in a theory of Einstein gravity with the quantum gravity cutoff-scale $L_*$, see also \cite{species}. If the black hole radius is much larger than $L_*$, the Hawking radiation is nearly thermal. One of the signals that the size of a black hole reached the scale $L_*$ is that its Hawking evaporation cannot maintain its thermality. The measure of thermality is given by a parameter $\dot{T} / T^2$, where $T$ is the Hawking temperature. It is easy to see that in an (approximately) thermal regime and for the emission of a single particle of spin $s \gg 1$ this parameter has the following form,
\begin{equation}  \label{Temp}
{\dot{T} \over T^2}  \sim  (2s + 1)  {1 \over S_{BH}} \,,
\end{equation}
where $S_{BH}$ is the black hole entropy \cite{Bek2}. Hence, the parameter becomes of order one once the black hole temperature reaches a critical value $T_* \sim M_P / \sqrt{s}$ and correspondingly the black hole reaches the size $L_* \sim \sqrt{s} L_P$. Therefore, one obtains the following bound on the gravitational cutoff and the spin of a particle
\begin{equation}  \label{Sstar}
L_*  \gtrsim \sqrt{s} L_P \,.
\end{equation}
One can straightforwardly generalize this to the presence of many massive particle species, for which $\sqrt{s} \rightarrow \sqrt{N} = \sqrt{\sum_j (2s_j + 1)}$, where $j$ runs over all massive species of spin $s_j$. The massless fields must also be included with the appropriate degeneracies. 

Furthermore, since the particle carries spin the remaining black hole will in general carry angular momentum after the Hawking radiation process involving the spin $s$ particle. This modifies the evaporation. In general, the maximal angular momentum of a black hole is obtained for a extremal Kerr black hole, at
\begin{align}
J_{\text{max}} \sim M^2 / M_P^2 \,.
\end{align}
Therefore, the smallest black hole that can accommodate an angular momentum of the size of the particle spin $s$ is of the size
\begin{align}
R \sim M / M_P^2 \sim \sqrt{J_{\text{max}}} / M_P = \sqrt{s} L_P \,,
\end{align}
which perfectly coincides with the quantum gravity cut-off scale \eqref{Sstar} derived above. This means that $L_*$ does not only indicate a scale at which the thermal behavior of the black hole evaporation breaks down, but also the scale beyond which an evaporation process involving the spin $s$ particle leads to a naked singularity within the classical gravitational approximation.\\

Finally, it is useful to compare the two bounds above. Although the intrinsically gravitational bound \eqref{Sstar} seems much more stringent than the non-gravitational one \eqref{Sbound}, it carries qualitatively different information. To see this, let us discuss some explicit numbers. Using the current experimental constraint $L_* \lesssim {\rm TeV}^{-1}$, the bound \eqref{Sstar} implies that a single elementary particle of mass $M \ll 1/L_*$ can have a spin as high as $s \sim 10^{32}$.  At the same time, such an elementary particle is excluded by the bound \eqref{Sbound}, which demands that an object with such a large spin cannot be an elementary particle in the usual sense and must acquire a notion of a new length-scale. This is necessary in order to make the localization much more energetically expensive than in the case of a Gaussian wave-packet of an elementary particle.\\

\section*{Acknowledgements}
We thank Cesar Gomez, Dieter L\"ust, Miguel Montero, Oriol Pujolas, Wilke Van Der Schee, John Stout, Stefan Vandoren, and Nico Wintergerst for discussions.  M.D.'s work is part of the D-ITP consortium, a program of the Netherlands Organisation for Scientific Research (NWO) that is funded by the Dutch Ministry of Education, Culture and Science (OCW). The work of G.D. was supported in part by the Humboldt Foundation under Humboldt Professorship Award, by the Deutsche Forschungsgemeinschaft (DFG, German Research Foundation) under Germany's Excellence Strategy - EXC-2111 - 390814868, and Germany's Excellence Strategy under Excellence Cluster Origins.


\begin{thebibliography}{10}

\bibitem{BekBound}
J.~D.~Bekenstein, {\em Universal upper bound on the entropy-to-energy ratio for bounded systems\/}, \href{http://dx.doi.org/10.1103/PhysRevD.23.287}{Phys. Rev. D {\bf 23} (1981) no.~2, 287--298}. 

\bibitem{BrBound}
H.~J.~Bremermann, {\em Quantum noise and information\/},  in {\em Proceedings of the Fifth Berkeley Symposium on Mathematical Statistics and Probability} {\bf 4} (1967), 15--20.

\bibitem{History}
D.~N.~Page, {\em The Bekenstein Bound}, \href{https://arxiv.org/abs/1804.10623}{\tt arXiv:1804.10623 [hep-th]}.\\
R.~M.~Wald, {\em Jacob Bekenstein and the Development of Black Hole Thermodynamics}, \href{https://arxiv.org/abs/1805.02302}{\tt arXiv:1805.02302 [physics.hist-ph]}.\\
R.~Bousso, {\em Black hole entropy and the Bekenstein bound}, \href{https://arxiv.org/abs/1810.01880}{\tt arXiv:1810.01880 [hep-th]}.

\bibitem{Gia}
G.~Dvali, {\em Area Law Saturation of Entropy Bound from Perturbative Unitarity in Renormalizable Theories}, \href{https://arxiv.org/abs/1906.03530}{\tt arXiv:1906.03530 [hep-th]}.\\
G.~Dvali, {\em Unitarity Entropy Bound: Solitons and Instantons}, \href{https://arxiv.org/abs/1907.07332}{\tt arXiv:1907.07332 [hep-th]}.

\bibitem{Bousso}
R.~Bousso, {\em Bound states and the Bekenstein bound}, \href{https://iopscience.iop.org/article/10.1088/1126-6708/2004/02/025}{JHEP {\bf 0402} (2004) 025}, \href{https://arxiv.org/abs/hep-th/0310148}{hep-th/0310148}.

\bibitem{species}
G.~Dvali, \textit{Black Holes and Large N Species Solution to the Hierarchy Problem}, \href{https://onlinelibrary.wiley.com/doi/abs/10.1002/prop.201000009}{Fortschr.\ Phys. {\bf 58} (2010), 528}, \href{https://arxiv.org/abs/arXiv:0706.2050}{\tt arXiv:0706.2050 [hep-th]}.\\
G.~Dvali and M.~Redi, \textit{Black Hole Bound on the Number of Species and Quantum Gravity at LHC}, \href{https://journals.aps.org/prd/abstract/10.1103/PhysRevD.77.045027}{Phys.\ Rev.  {\bf D77} (2008),  045027}, \href{https://arxiv.org/abs/arXiv:0710.4344}{\tt arXiv:0710.4344 [hep-th]}.\\
G.~Dvali and C.~Gomez, \textit{Quantum Information and Gravity Cutoff in Theories with Species}, \href{https://www.sciencedirect.com/science/article/pii/S0370269309003025?via\%3Dihub}{Phys.\ Lett.  {\bf B674} (2009),  303}, \href{https://arxiv.org/abs/arXiv:0812.1940}{\tt arXiv:0812.1940 [hep-th]}.

\bibitem{Bek2}
J.~D.~Bekenstein, {\em Black holes and entropy\/}, \href{http://dx.doi.org/10.1103/PhysRevD.7.2333}{Phys. Rev. D {\bf 7} (1973) no.~8, 2333--2346}.
 
\end{thebibliography}
\end{document}